\documentclass[10pt,twocolumn,letterpaper]{article}

\usepackage{iccv}
\usepackage{times}
\usepackage{epsfig}
\usepackage{graphicx}
\usepackage{amsmath}
\usepackage{amssymb}
\usepackage{cuted} 
\usepackage{capt-of}
\usepackage[font={small}]{caption}
\usepackage{afterpage}
\usepackage{amsmath}
\usepackage{enumitem}
\usepackage{booktabs} 
\usepackage{multirow} 
\usepackage{subcaption} 
\usepackage{url} 

\usepackage[usenames]{color}
\usepackage[usenames,dvipsnames]{xcolor}

\usepackage[breaklinks=true,bookmarks=false]{hyperref}

\iccvfinalcopy

\ificcvfinal\fi

\begin{document}

\title{Everybody Dance Now}

\author{Caroline Chan\thanks{C. Chan is currently a graduate student at MIT CSAIL.}
\qquad
Shiry Ginosar
\qquad
Tinghui Zhou\thanks{T. Zhou is currently affiliated with Humen, Inc.}
\qquad
Alexei A. Efros \\
\\
UC Berkeley
}

\maketitle
\ificcvfinal\thispagestyle{empty}\fi

\begin{strip}\centering
\vspace{-0.6in}
\includegraphics[width=\textwidth]{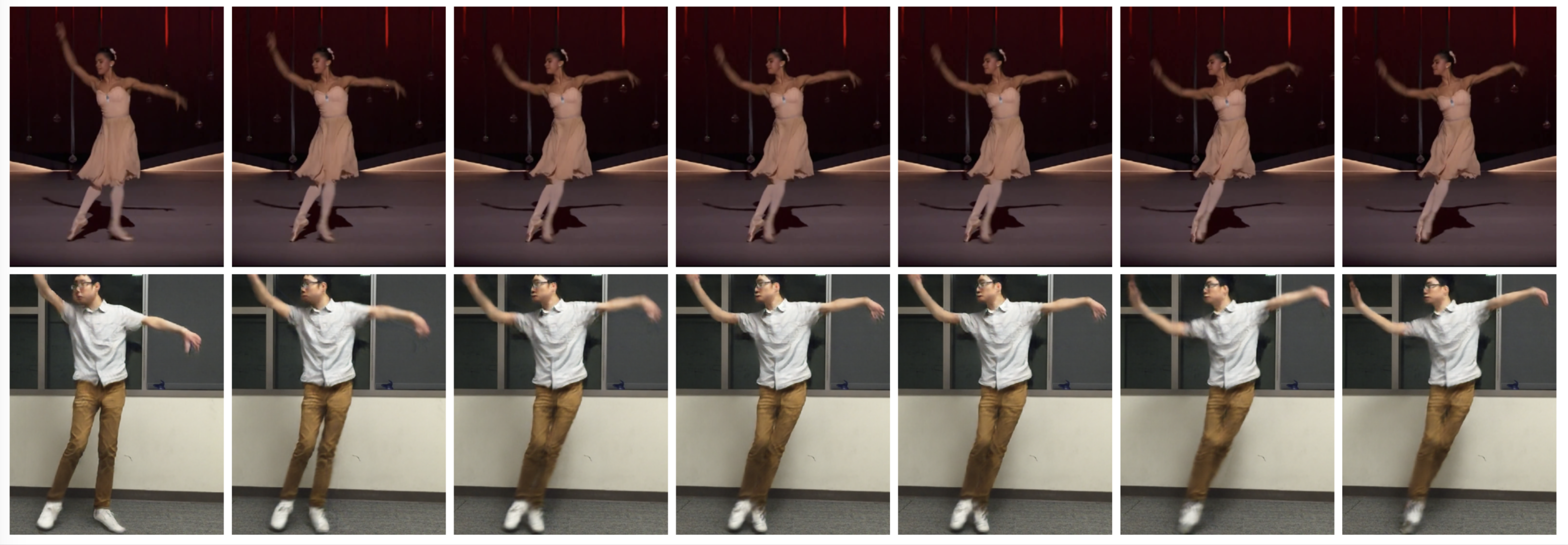}
\vspace{-.3in}
\captionof{figure}{{\bf``Do as I Do'' motion transfer:}  given a YouTube clip of a ballerina (top), and a video of a graduate student performing various motions, our method transfers the ballerina's performance onto the student (bottom). Video: \small{\url{https://youtu.be/mSaIrz8lM1U}}}
\label{fig:feature-graphic}
\end{strip}
\begin{abstract}

   This paper presents a simple method for ``do as I do" motion transfer: given a source video of a person dancing, we can transfer that performance to a novel (amateur) target after only a few minutes of the target subject performing standard moves. We approach this problem as video-to-video translation using pose as an intermediate representation. To transfer the motion, we extract poses from the source subject and apply the learned pose-to-appearance mapping to generate the target subject. We predict two consecutive frames for temporally coherent video results and introduce a separate pipeline for realistic face synthesis. Although our method is quite simple, it produces surprisingly compelling results (see video). This motivates us to also provide a forensics tool for reliable synthetic content detection, which is able to distinguish videos synthesized by our system from real data. In addition, we release a first-of-its-kind open-source dataset of videos that can be legally used for training and motion transfer.
\end{abstract}

\section{Introduction}

Consider the two video sequences on Figure~\ref{fig:feature-graphic}. The top
row is the input -- it is a YouTube clip of a ballerina (the
{\em source} subject) performing a sequence of motions. The bottom row
is the output of our algorithm. It corresponds to frames of a different
person (the {\em target} subject) apparently performing the same motions. The twist is that the target person never performed the same exact sequence
of motions as the source, and, indeed, knows nothing about ballet. He was instead filmed performing a set of standard moves,
without specific reference to the precise actions of the
source. 
And, as is obvious from the figure,
the source and the target are of different genders, have different builds, and wear different clothing.

In this work, we propose a simple but surprisingly effective approach for ``Do as I Do" video retargeting -- automatically transferring the motion from a source to a target subject.
Given two videos -- one of a \textit{target} person whose appearance we wish to synthesize, and the other of a \textit{source} subject whose motion we wish to impose onto our target person -- we transfer motion between these subjects by learning a simple video-to-video translation. 
With our framework, we create a variety of videos, enabling untrained amateurs to spin and twirl like ballerinas, perform martial arts kicks, or dance as vibrantly as pop stars.

To transfer motion between two video subjects in a frame-by-frame manner, we must learn a mapping between images of the two individuals. Our goal is, therefore, to discover an image-to-image translation~\cite{isola2016image} between the source and target sets. However, we do not have corresponding pairs of images of the two subjects performing the same motions to supervise learning this translation. Even if both subjects perform the same routine, it is still unlikely to have an exact frame to frame pose correspondence due to body shape and motion style unique to each subject.

\begin{figure}
  \includegraphics[width=\linewidth]{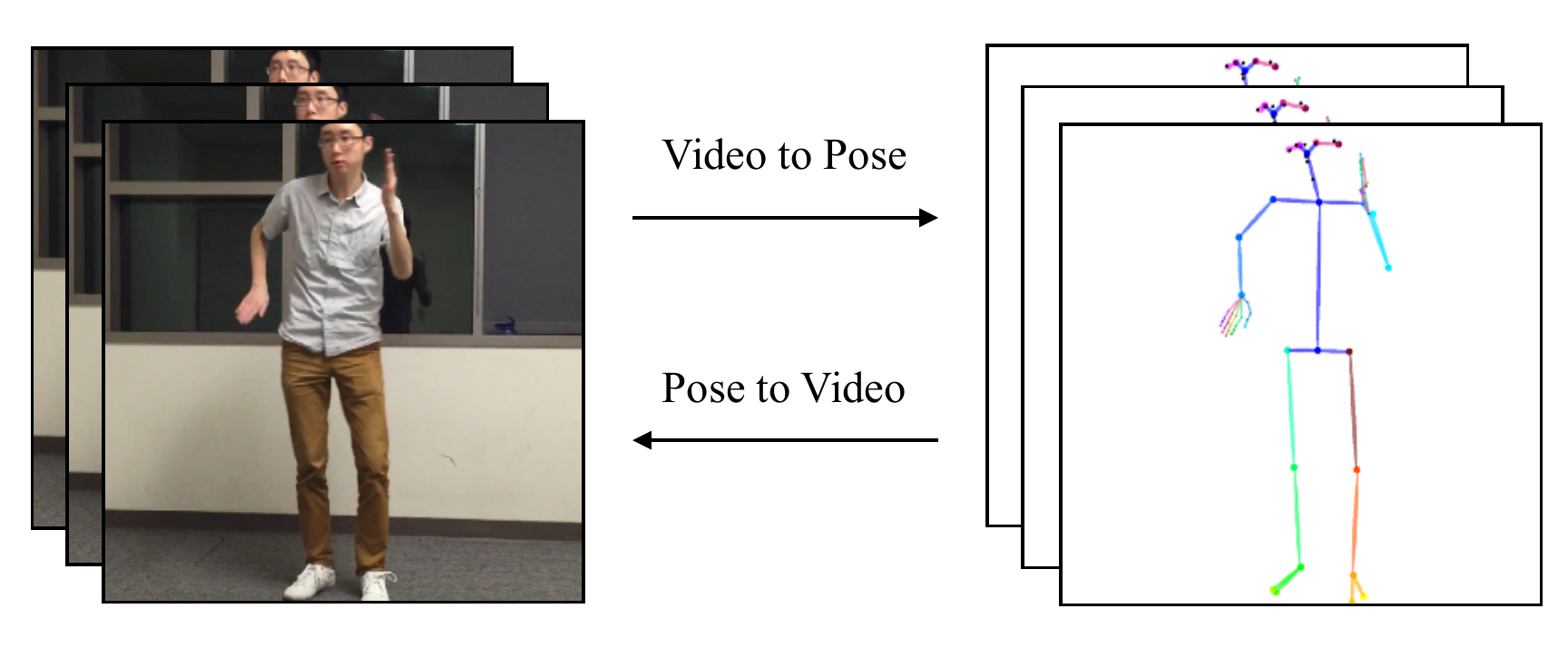}
  \vspace{-.3in}
  \caption{Our method creates correspondences by detecting poses in video frames (Video to Pose) and then learns to generate images of the target subject from the estimated pose (Pose to Video).}
  \label{fig:corres}
  \vspace{-.2in}
\end{figure}

We observe that keypoint-based pose preserves motion signatures over time while abstracting away as much subject identity as possible and can serve as an intermediate representation between any two subjects. We therefore use pose stick figures obtained from off-the-shelf human pose detectors, such as OpenPose~\cite{cao2017realtime,simon2017hand,wei2016cpm}, as an intermediate representation for frame-to-frame transfer, as shown in Figure~\ref{fig:corres}. We then learn an image-to-image translation model between pose stick figures and images of our target person. To transfer motion from source to target, we input the pose stick figures from the source into the trained model to obtain images of the target subject in the same pose as the source. 

The central contribution of our work is a surprisingly simple method for generating compelling results on human motion transfer. We demonstrate complex motion transfer from realistic in-the-wild input videos and synthesize high-quality and detailed outputs (see Section~\ref{sec:results} and our video for examples). Motivated by the high quality of our results, we introduce an application for detecting if a video is real or synthesized by our method. We strongly believe that it is important for work in image synthesis to explicitly address the issue of fake detection (Section~\ref{sec:fake}).

Furthermore, we release a two-part dataset: First, five long single-dancer videos which we filmed ourselves that can be used to train and evaluate our model, and second, a large collection of short YouTube videos that can be used for transfer and fake detection. We specifically designate the single-dancer data to be high-resolution open-source data for training motion transfer and video generation methods. The subjects whose data we release have all consented to allowing the data to be used for research purposes. For more details, see our project website \url{https://carolineec.github.io/everybody_dance_now} .

\begin{figure*}
\centering
  \includegraphics[width=0.90\linewidth]{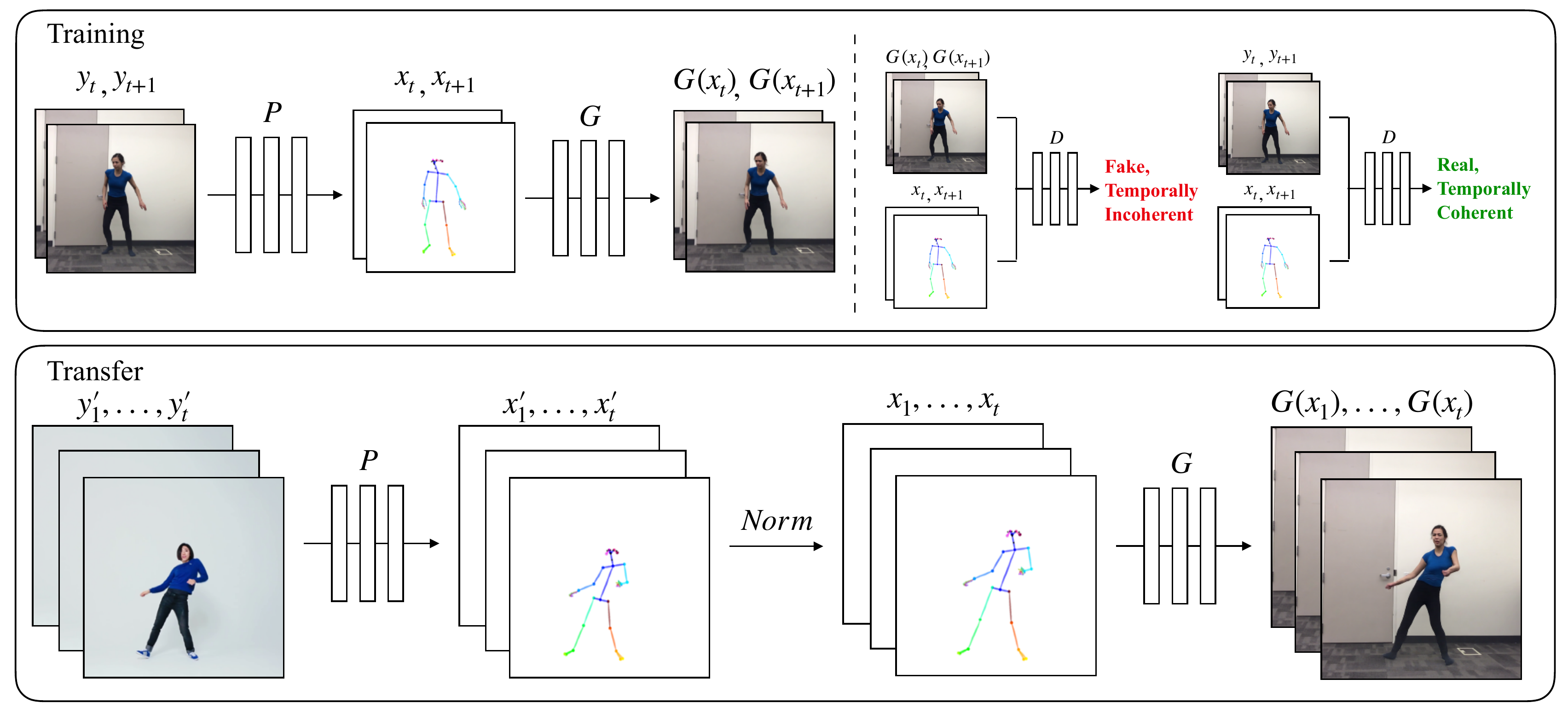}
  \vspace{-.1in}
    \caption{ (Top) \textbf{Training}: Our model uses a pose detector $P$ to create pose stick figures from video frames of the target subject. We learn the mapping $G$ alongside an adversarial discriminator $D$ which attempts to distinguish between the ``real'' correspondences $(x_t, x_{t+1}), (y_t, y_{t+1})$ and the ``fake'' sequence $(x_t, x_{t+1}), (G(x_t), G(x_{t+1}))$ . (Bottom) \textbf{Transfer}: We use a pose detector $P$ to obtain pose joints for the source person that are transformed by our normalization process $Norm$ into joints for the target person for which pose stick figures are created. Then we apply the trained mapping $G$.}
  \label{fig:system}
  \vspace{-0.2in}
\end{figure*}

\section{Related Work}

Over the last two decades there has been extensive work dedicated to motion transfer. Early methods focused on creating new content by manipulating existing video footage~\cite{bregler1997video,Efros03,MoriBEEM04}. For example, Video Rewrite~\cite{bregler1997video} creates videos of a subject saying a phrase they did not originally utter by finding frames where the mouth position matches the desired speech. Efros et al.~\cite{Efros03} use optical flow as a descriptor to match different subjects performing similar actions allowing ``Do as I do'' and ``Do as I say'' retargeting. Classic computer graphics approaches to motion transfer attempt to perform this in 3D.  Ever since the retargeting problem was proposed between animated characters~\cite{gleicher1998retargetting}, solutions have included the use of inverse kinematic solvers~\cite{lee1999hierarchical} and retargeting between significantly different 3D skeletons~\cite{hecker2008real}. Our approach is similarly designed for in-the-wild video subjects, although we learn to synthesize novel motions rather than manipulating existing frames and we use 2D representations.

Several approaches rely on calibrated multi-camera setups to `scan' a target actor and manipulate their motions in a new video through a fitted 3D model of the target. To obtain 3D information, Cheung et al.~\cite{cheung2004markerless} propose an elaborate multi-view system to calibrate a personalized kinematic model, obtain 3D joint estimations, and render images of a human subject performing new motions.
Xu et al.~\cite{xu2011video} use multi-view captures of a target subject performing simple motions to create a database of images and transfer motion through a fitted 3D skeleton and corresponding surface mesh for the target. Work by Casas et al. use 4D Video Textures~\cite{casasEG2014} to compactly store a layered texture representation of a scanned target person and use their temporally coherent mesh and data representation to render video of the target subject performing novel motions. In contrast, our approach explores motion transfer between 2D video subjects and avoid data calibration and lifting into 3D space.

Similarly to our method, recent works have applied deep learning for reanimation in different applications and rely on more detailed input representations. Given synthetic renderings, an interior face model, and a gaze map as input, Kim et al.~\cite{kim2018DeepVideo} transfer head position and facial expressions between human subjects and render their results in detailed portrait videos. Our problem is analogous to this work except we retarget full body motion, and the inputs to our model as 2D pose stick figures as opposed to more detailed 3D representations.  Similarly, Martin-Brualla et al.~\cite{Martin-Brualla:2018:LEP:3272127.3275099} apply neural re-rendering to enhance rendering of human motion capture for VR/AR purposes. The primary focus of this work is to render realistic humans in real time and similarly uses a deep network to synthesize their final result, but unlike our work does not address motion transfer between subjects. Villegas et al.~\cite{villegas2018neural} focus on retargeting motion between rigged skeletons and demonstrate reanimation in 3D characters without supervised data. Similarly, we learn to retarget motion using a skeleton-like intermediate representation, however we transfer full body motion between human subjects who are not rigged to the skeleton unlike animated characters.

Recent methods focus on disentangling motion from appearance and synthesizing videos with novel motion~\cite{tulyakov2017mocogan,baddar2017dynamics}. MoCoGAN~\cite{tulyakov2017mocogan} employs unsupervised adversarial training to learn this separation and generates videos of subjects performing novel motions or facial expressions. This theme is continued in Dynamics Transfer GAN~\cite{baddar2017dynamics} which transfers facial expressions from a source subject in a video onto a target person given in a static image. Similarly, we apply our representation of motion to different target subjects to generate new motions. However, in contrast to these methods we specialize on synthesizing detailed dance videos. 

Modern approaches have shown success in generating detailed single images of human subjects in new poses~\cite{balakrishnan2018synthesizing, de2018semi, de2019conditional, joo2018generating,lassner2017generative, ma2017pose,ma2018disentangled,Siarohin_2018_CVPR,villegas2017learning,vunet2018,zanfir2018human}. Works including Ma et al.~\cite{ma2017pose, ma2018disentangled} and Siarohin et al.~\cite{Siarohin_2018_CVPR} have introduced novel architectures and losses for this purpose. Furthermore, ~\cite{pos_iccv2017, villegas2017learning} have shown pose is an effective supervisory signal for future prediction and video generation. However these works are not designed specifically for motion transfer. Rather than generating possible views of a previously unseen person from a single input image, we are interested in learning the style of a single, known person from large amounts of personalized video data and synthesizing them dancing in a detailed high-resolution video.

Concurrent with our work,~\cite{aberman2019deep, Recycle-GAN, Liu2018Neural, wang2018vid2vid} learn mappings between videos and demonstrate motion transfer between faces and from poses to body. Wang et al.~\cite{wang2018vid2vid} achieves results of similar quality to ours with a more complex method and significantly more computational resources.

Our work is made possible by recent rapid advances along two separate directions: robust pose estimation, and realistic image-to-image translation. Modern pose detection systems including OpenPose~\cite{cao2017realtime, simon2017hand, wei2016cpm} and DensePose~\cite{Guler2018DensePose} allow for surprisingly reliable and fast pose extraction in a variety of scenarios. At the same time, the recent emergence of image-to-image translation models, pix2pix~\cite{isola2016image}, CoGAN~\cite{liu2016coupled}, UNIT~\cite{liu2017unsupervised},  CycleGAN~\cite{CycleGAN2017}, 
DiscoGAN~\cite{kim2017learning}, Cascaded Refinement Networks~\cite{chen2017photographic}, and pix2pixHD~\cite{wang2017highres},  have enabled high-quality single-image generation. We build upon these two building blocks by using pose detection as an intermediate representation and extending upon single-image generation to synthesize temporally-coherent, surprisingly realistic videos.

\section{Method}
Given a video of a source person and another of a target person, our goal is to generate a new video of the target enacting the same motions as the source. To accomplish this task, we divide our pipeline into three stages -- pose detection, global pose normalization, and mapping from normalized pose stick figures to the target subject. See Figure~\ref{fig:system} for an overview of our pipeline. In the pose detection stage we use a pre-trained state-of-the-art pose detector to create pose stick figures given frames from the source video. The global pose normalization stage accounts for differences between the source and target body shapes and locations within the frame. Finally, we design a system to learn the mapping from the pose stick figures to images of the target person using adversarial training. Next we describe each stage of our system.

\subsection{Pose Encoding and Normalization}

\paragraph{Encoding body poses}
\label{pose_estimation}
To encode the body pose of a subject image, we use a pre-trained pose detector $P$ (OpenPose~\cite{cao2017realtime,simon2017hand,wei2016cpm}) which accurately estimates 2D $x,y$ joint coordinates. We then create a colored pose stick figure by plotting the keypoints and drawing lines between connected joints as shown in Figure~\ref{fig:corres}.
\vspace{-0.15in}
\paragraph{Global pose normalization} \label{posenorm}
In different videos, subjects may have different limb proportions or stand closer or farther to the camera than one another. Therefore when retargeting motion between two subjects, it may be necessary to transform the pose keypoints of the source person so that they appear in accordance with the target person's body shape and location as in the \textbf{Transfer} section of Figure~\ref{fig:system}. We find this transformation by analyzing the heights and ankle positions for the poses of each subject and use a linear mapping between the closest and farthest ankle positions in both videos. After gathering these positions, we calculate the scale and translation for each frame based on its corresponding pose detection. Details of this process are described in Section~\ref{posenorm}.

\subsection{Pose to Video Translation}

Our video synthesis method is based off of an adversarial single frame generation process presented by Wang et al.~\cite{wang2017highres}. In the original conditional GAN setup, the generator network $G$ engages in a minimax game against multi-scale discriminator $D = (D_1, D_2, D_3)$. The generator must synthesize images in order to fool the discriminator which must discern between ``real" (ground truth) images and ``fake'' images produced by the generator. The two networks are trained simultaneously and drive each other to improve - $G$r learns to synthesize more detailed images to deceive $D$ which in turn learns differences between generated outputs and ground truth data. For our purposes, $G$ synthesizes images of a person given a pose stick figure.

Such single-frame image-to-image translation methods are not suitable for video synthesis as they produce temporal artifacts and cannot generate the fine details important in perceiving humans in motion. We therefore add a learned model of temporal coherence as well as a module for high resolution face generation.

\begin{figure}
  \includegraphics[width=\linewidth]{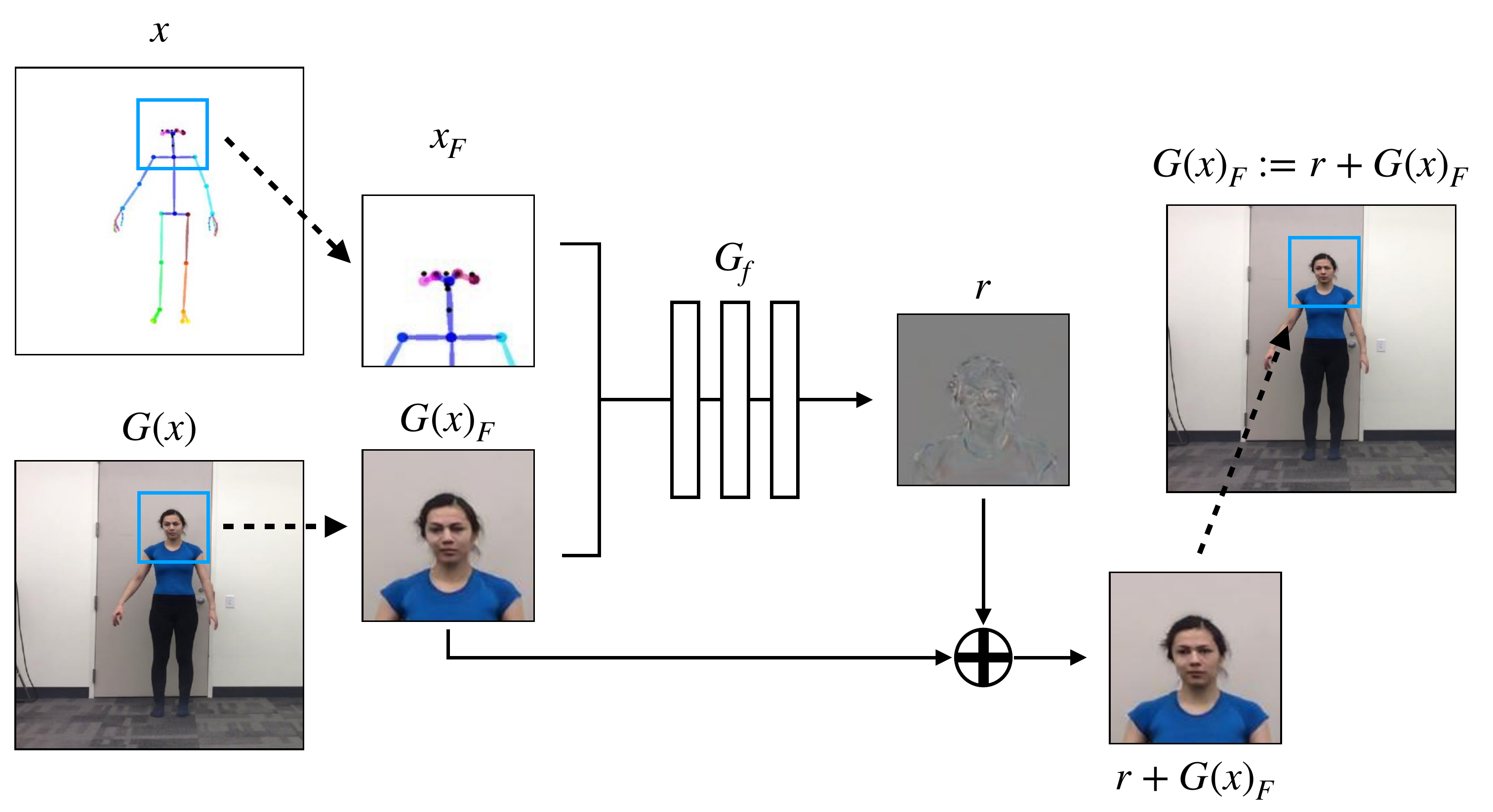}
    \vspace{-.3in}
  \caption{Face GAN setup. Residual is predicted by generator $G_f$ and added to the original face prediction from the main generator.}
  \vspace{-0.2in}
  \label{fig:facegan}
\end{figure}

\paragraph{Temporal smoothing}
To create video sequences, we modify the single image generation setup to enforce temporal coherence between adjacent frames as shown in Figure~\ref{fig:system} (top right). Instead of generating individual frames, we predict two consecutive frames where the first output $G(x_{t-1})$ is conditioned on its corresponding pose stick figure $x_{t-1}$ and a zero image $z$ (a placeholder since there is no previously generated frame at time $t-2$). The second output $G(x_{t})$ is conditioned on its corresponding pose stick figure $x_{t}$ and the first output $G(x_{t-1})$.
Consequently, the discriminator is now tasked with determining both the difference in realism and temporal coherence between the ``fake'' sequence $(x_{t-1}, x_{t}, G(x_{t-1}), G(x_{t}))$ and ``real'' sequence $(x_{t-1}, x_{t}, y_{t-1}, y_{t})$. The temporal smoothing changes are now reflected in the updated GAN objective
\begin{eqnarray}
& \mathcal{L}_{\mathrm{smooth}}(G,D) = \mathbb{E}_{(x,y)} [\log D(x_{t}, x_{t+1} ,y_{t}, y_{t+1})] & \nonumber \\
& + \mathbb{E}_x [\log(1 - D(x_{t}, x_{t+1}, G(x_{t}), G(x_{t+1}))] &
\end{eqnarray}
\paragraph{Face GAN} \vspace{-.1in}
We add a specialized GAN setup to add more detail and realism to the face region as shown in Figure~\ref{fig:facegan}.
After generating the full image of the scene with the main generator $G$, we input a smaller section of the image centered around the face (i.e. $128\times128$ patch centered around the nose keypoint), $G(x)_F$, and the input pose stick figure sectioned in the same fashion, $x_F$, to another generator $G_f$ which outputs a residual $r = G_f(x_F, G(x)_F)$. The final synthesized face region is the addition of the residual with the face region of the main generator $r + G(x)_F$. A discriminator $D_f$ then attempts to discern the ``real" face pairs $(x_F, y_F)$ from the ``fake" face pairs $(x_F, r + G(x)_F)$, similarly to the original pix2pix~\cite{isola2016image} objective:
\begin{eqnarray}
& \mathcal{L}_{\mathrm{face}}(G_f, D_f) = \mathbb{E}_{(x_F,y_F)} [\log D_f(x_F,y_F)] & \nonumber \\
& + \mathbb{E}_{x_F} [\log \big( 1 - D_f(x_F, G(x)_F + r) \big) ] . &
\end{eqnarray}
Here $x_F$ is the face region of the original pose stick figure $x$ and $y_F$ is the face region of ground truth target person image $y$. Similarly to the full image, we add a perceptual reconstruction loss on comparing the final face $r + G(x)_F$ to the ground truth target person's face $y_F$.

\begin{figure*}
  \includegraphics[width=\linewidth]{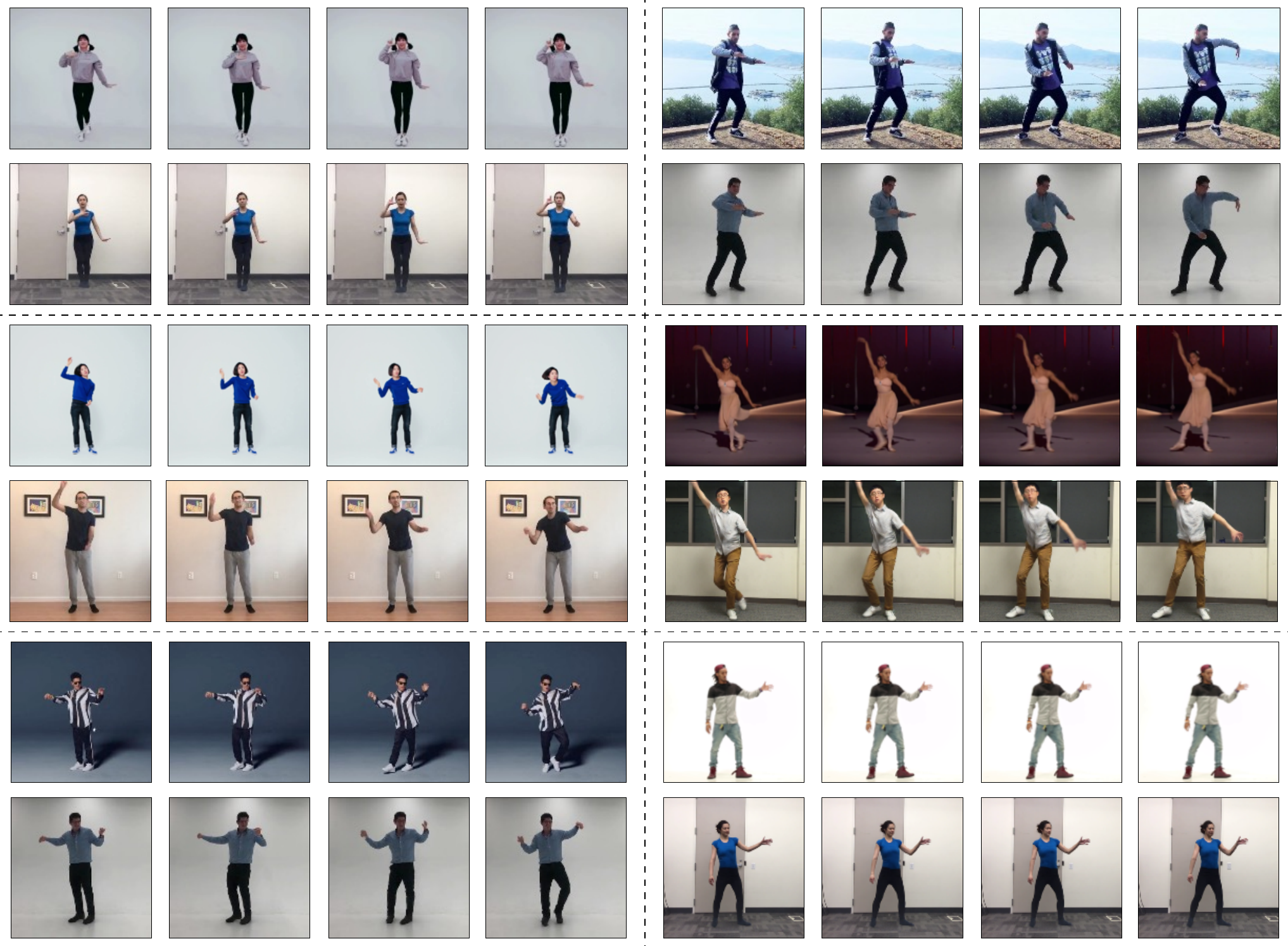}
    \caption{Transfer results. In each section we show four consecutive frames. The top row shows the source subject and the bottom row shows the synthesized outputs of the target person.}
  \label{fig:results}
  \vspace{-0.1in}
\end{figure*}

\subsection{Full Objective}
We employ training in stages where the full image GAN is optimized separately from the specialized face GAN. First we train the main generator and discriminator $(G, D)$ during which the full objective is -
\begin{eqnarray} 
& \min\limits_{G} (( \max\limits_{\small{D_i}} \textstyle\sum\limits_{\small{k_i}} \mathcal{L}_{\mathrm{smooth}}(G, D_k))
+ \lambda_{FM} \textstyle\sum\limits_{\small{k_i}} \mathcal{L}_{\mathrm{FM}}(G, D_k) \nonumber & \\
& + \lambda_{P} (\mathcal{L}_{P}(G(x_{t-1}),y_{t-1})
+ \mathcal{L}_{P}(G(x_{t}),y_{t}))) &
\end{eqnarray}
Where $i={1,2,3}$.
Here, $\mathcal{L}_{\mathrm{GAN}}(G, D)$ is the single image adversarial loss presented in the original pix2pix paper~\cite{isola2016image}:
\begin{equation}
\mathcal{L}_{\mathrm{GAN}}(G, D) = \mathbb{E}_{(x,y)} [\log D(x,y)] + \mathbb{E}_x [\log(1 - D(x, G(x))]
\end{equation}
$\mathcal{L}_{\mathrm{FM}}(G, D)$ is the discriminator feature-matching loss presented in pix2pixHD, and $\mathcal{L}_{P}(G(x), y)$ is the perceptual reconstruction loss~\cite{Johnson2016Perceptual} which compares pretrained VGGNet~\cite{simonyan2014very} features at different layers of the network (fully specified in the Section~\ref{implementation}).

After this stage, the full image GAN weights are frozen and we optimize the face GAN with objective
\begin{eqnarray} 
\min\limits_{G_f} \bigg( \Big( \max\limits_{D_f} \mathcal{L}_{\mathrm{face}}(G_f, D_f) \Big) + \lambda_{P} \mathcal{L}_{P}(r + G(x)_F, y_F) \bigg)
\end{eqnarray}
where $\mathcal{L}_{\mathrm{FM}}(G, D)$ is the discriminator feature-matching loss presented in pix2pixHD, and $\mathcal{L}_{P}$ is a perceptual reconstruction loss~\cite{Johnson2016Perceptual} which compares pretrained VGGNet~\cite{simonyan2014very} features at different layers of the network. For training details see Section~\ref{implementation}.

\section{Experiments} 

We compare our performance to baseline methods on multiple target subjects and source motions.

\subsection{Setup}
\label{sec:setup}

We collect two types of data long, open-source, single-dancer \emph{target} videos which we film ourselves to train our model on and make publicly available, and in-the-wild \emph{source} videos collected online for motion transfer. The filming set-up for target videos and collection method for source videos are detailed in Section~\ref{collection}.

\vspace{-0.1in}
\paragraph{Baseline methods} 1) \textbf{Nearest Neighbors}. For each source video frame, we retrieve the closest match in the training target sequence using the following pose distance metric: For two poses $p, p'$ each with $n$ joints $p_1, ..., p_n$ and $p'_1,..., p'_n$, we define the distance between them as the normalized sum of the L2 distances between the corresponding joints $p_k = (x_k, y_k)$ and $p'_k = (x'_k, y'_k)$:
\begin{equation}
d(p, p') = \frac{1}{n} \sum_{k=1}^{n} \lVert p_k - p'_k \rVert_2
\label{metric}
\vspace{-0.1in}
\end{equation}
The adjacent target matches frames are then concatenated into a frame-by-frame nearest neighbors sequence.

\noindent
2) {\bf Balakrishnan \etal (PoseWarp)}~\cite{balakrishnan2018synthesizing} generate images of a given target subject in a new pose. While, unlike ours, this method is designed for single image synthesis, we use it to synthesize a video frame-by-frame for comparison.

\vspace{-0.1in}
\paragraph{Ablation conditions} 1) \textbf{Frame-by-frame synthesis} (FBF). In this condition we ablate our temporal smoothing setup and apply pix2pixHD~\cite{wang2017highres} on a per-frame basis. 2) \textbf{Temporal smoothing} (FBF+TS). In this condition we ablate the Face GAN module to study the difference it makes on the final result. 3) \textbf{Our model} (FBF+TS+FG). uses both temporal smoothing and a Face GAN.

\vspace{-0.1in}
\paragraph{Evaluation metrics} We use perceptual studies on Mechanical Turk for evaluating the video results of our final method in comparison to ablated conditions and baselines. For the ablation study, we further measure the quality of each synthesized frame using two metrics: 1) \textbf{SSIM}. Structural Similarity~\cite{wang2004image} and 2) \textbf{LPIPS} Learned Perceptual Image Patch Similarity~\cite{zhang2018perceptual}.
We examined the pose distance seen in Equation~\ref{metric} to measure the similarity between input and synthesized pose. However, we found this ‘distance’ to be not very informative due to noisy detections. 

\subsection{Quantitative Evaluation}
\label{quantitative}
We quantitatively compare our approach against the baselines, and then against ablated versions of our method. 
\subsubsection{Comparison to Baselines}
\label{quantitative-baselines}

\begin{table}
\centering
\setlength{\tabcolsep}{2pt}
\begin{tabular}{lcccccc}
Method & $1$ & $2$ & $3$ & $4$ & $5$ & Total\\
\midrule
NN & $95.9\%$ & $96.4\%$ & $94.6\%$ & $95.8\%$ & $94.7\%$  & $95.1\%$\\
PoseWarp~\cite{balakrishnan2018synthesizing} & $83.1\%$ & $69.9\%$ & $88.7\%$ & $84.6\%$ & $74.4\%$ & $83.3\%$\\
\end{tabular}
\caption{Comparison to baselines using perceptual studies for subjects $1$ through $5$ and in total average. We report the percentage of time participants chose \textbf{our} method as more realistic than the baseline.}
\label{table:mturk-baselines}
\vspace{-0.1in}
\end{table}

\begin{table}
\centering
\setlength{\tabcolsep}{2pt}
\begin{tabular}{lcccccc}
Method & $1$ & $2$ & $3$ & $4$ & $5$ & Total\\
\midrule
NN & $85\%$ & $93\%$ & $94\%$ & $90\%$ & $91\%$  & $91.2\%$\\
PoseWarp [3] & $77.5\%$ & $70\%$ & $80\%$ & $90\%$ & $78.7\%$ & $79.1\%$\\
\end{tabular}
\caption{Comparison of our method without Face GAN (FBF+TS variant) to baselines for subjects $1$ through $5$ and in total average. We report the percentage of time participants chose the FBF+TS ablation as more realistic than the baseline.}
\label{table:mturk-baselines-noface}
\vspace{-0.2in}
\end{table}

We compare our method to baselines on the same transfer task for all subjects for which we filmed longer videos. From a single out-of-sample source video, we synthesize a transfer video for every baseline-subject pair. We then crop the same $10$-second snippets of video for each baseline and subject pair and use these for our perceptual studies.

Participants on MTurk watched a series of video pairs. In each pair, one video was synthesized using our method; the other by a baseline. They were then asked to pick the more realistic one. Videos of resolution $144\times 256$ (as this is the highest resolution that PoseWarp baseline can produce) were shown, and after each pair, participants were given unlimited time to respond. Each task consisted of $18$ pairs of videos and was performed by $100$ distinct participants. Table~\ref{table:mturk-baselines} displays the results of this study and shows that participants indicated our method is more realistic $95.1\%$ and $83.3\%$ of the time on average in comparison to the Nearest Neighbors and PoseWarp~\cite{balakrishnan2018synthesizing} baselines respectively.

We include an additional perceptual study to verify our method is not preferred over the others simply due to more emphasis on face synthesis. We compare the FBF+TS variant (without the Face GAN module) to both baselines in Table~\ref{table:mturk-baselines-noface}. We find that the FBF+TS ablation is consistently preferred, albeit slightly less than our full model, over the Nearest Neighbors and PoseWarp baselines $91.2\%$ and $79.1\%$ of the time on average respectively.

\subsubsection{Ablation Study}
\label{ablation}

\begin{table}
\centering
\begin{subtable}[c]{0.48\textwidth}
\setlength{\tabcolsep}{5pt}
\begin{tabular}{clccc}
\multicolumn{1}{c}{Region}& \multicolumn{1}{c}{Metric} & FBF & FBF+TS & FBF+TS+FG \\
\midrule
\parbox[t]{1mm}{\multirow{2}{*}{\rotatebox[origin=c]{90}{Face}}}
& SSIM & $0.784$ & $0.811$ & $\mathbf{0.816}$ \\
& LPIPS & $0.045$ & $0.039$ & $\mathbf{0.036}$ \\
\midrule
\parbox[t]{1mm}{\multirow{2}{*}{\rotatebox[origin=c]{90}{Body}}}
& SSIM & $0.828$ & $\mathbf{0.838}$ & $\mathbf{0.838}$ \\
& LPIPS & $0.057$ & $0.051$ & $\mathbf{0.050}$ \\
\end{tabular}
\caption{Metric comparison for synthesized face (top) and full-body (bottom) regions. Metrics are averaged over the $5$ subjects. For SSIM higher is better. For LPIPS lower is better.}
\label{table:ablation-all}
\end{subtable}

\vspace{0.1in}

\begin{subtable}[c]{0.48\textwidth}
\centering
\setlength{\tabcolsep}{2pt}
\begin{tabular}{lcccccc}
Condition & $1$ & $2$ & $3$ & $4$ & $5$ & Total\\
\midrule
FBF & $54.1\%$ & $69.7\%$ & $62.4\%$ & $53.8\%$ & $60.0\%$  & $58.8\%$\\
FBF+TS & $59.6\%$ & $56.4\%$ & $50.3\%$ & $53.0\%$ & $53.1\%$ & $53.9\%$\\
\end{tabular}
\caption{Perceptual study results for subjects $1$ through $5$ and in total average. We report the percentage of time participants chose \textbf{our} method as more realistic than the ablated conditions.}
\label{table:ablation-mturk}
\end{subtable}
\caption{Ablation studies. We compare frame-by-frame synthesis (FBF), adding temporal smoothing (FBF+TS) and our final model with temporal smoothing and Face GAN modules (FBF+TS+FG).}
\vspace{-0.1in}
\end{table}

\begin{table}
\centering
\setlength{\tabcolsep}{2pt}
\begin{tabular}{lcccccc}
Condition & $1$ & $2$ & $3$ & $4$ & $5$ & Total\\
\midrule
Prefer FBF+TS & $60.5\%$ & $62\%$ & $57.5\%$ & $50\%$ & $62.5\%$  & $58.5\%$\\
\end{tabular}
\caption{Comparison of our method without Face GAN (FBF+TS) to the FBF ablation for subjects $1$ through $5$ and in total average. We report the percentage of time participants chose the FBF+TS ablation over the FBF ablation.}
\label{table:ablation-mturk-noface}
\vspace{-0.1in}
\end{table}

We perform an ablation study on held-out test data of the target subject (the source and target are the same) since we do not have paired same-pose frames across subjects.

As shown in Table~\ref{table:ablation-all}(bottom), both SSIM and LPIPS scores are similar for all model variations on the body regions. Scores on full images are even more similar, as the ablated models have no difficulty generating the static background. However, Table~\ref{table:ablation-all}(top) demonstrates the effectiveness of our face residual generator by showing the improvement of our full model over the the FBF+TS condition.

As these comparisons are in a frame-by-frame fashion they do not emphasize the usefulness of our temporal smoothing setup. The effect of this module can be seen in the qualitative video results and in the perceptual studies results in Table~\ref{table:ablation-mturk}. Here we see that our method is preferred $58.8\%$ and $53.3\%$ of the time over frame-by-frame synthesis and the No Face GAN (FBF+TS) setup respectively. In general, this shows that incorporating temporal information at training time positively influences video results. Although the effect of the Face GAN can be be somewhat subtle, overall this addition benefits our results, especially in the case of subject $1$ whose training video is very sharp where facial details are easily visible.

\begin{figure}
\centering
  \includegraphics[width=\linewidth]{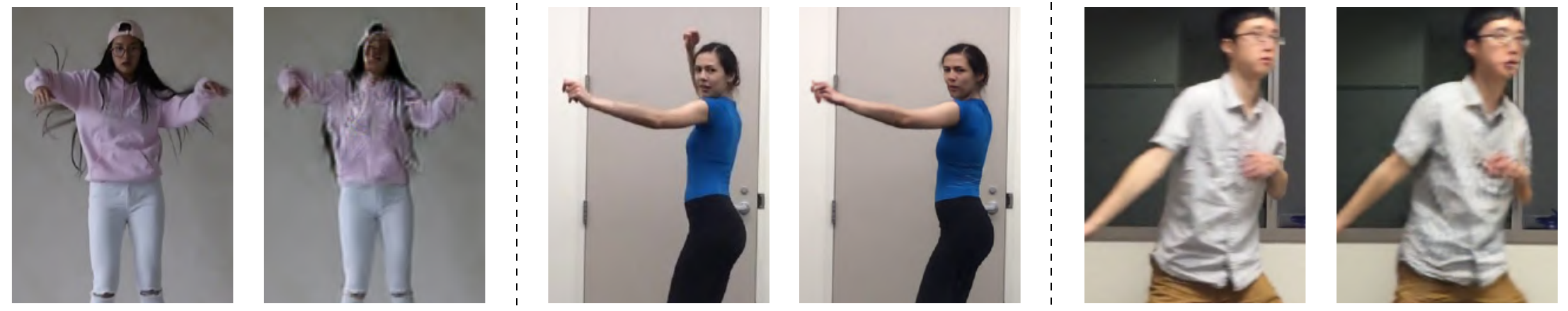}
  \caption{Failure cases. Ground truth appearance reference (left) followed by our results (right).}
  \vspace{-0.25in}
\label{fig:failure-cases}
\end{figure}

We further compare our method without the Face GAN (FBF+TS) to the frame-by-frame (FBF) ablation to verify our temporal smoothing setup alone improves result quality. Table~\ref{table:ablation-mturk-noface} reports that the FBF+TS ablation is preferred on average over the FBF alone. Note that for subject $4$ FBF produced noticeable flickering, but FBF+TS introduced texture artifacts on his loose shirt (see Figure~\ref{fig:failure-cases}).

\subsection{Qualitative Results}
\label{sec:results}
\begin{figure}
\centering
  \includegraphics[width=\linewidth]{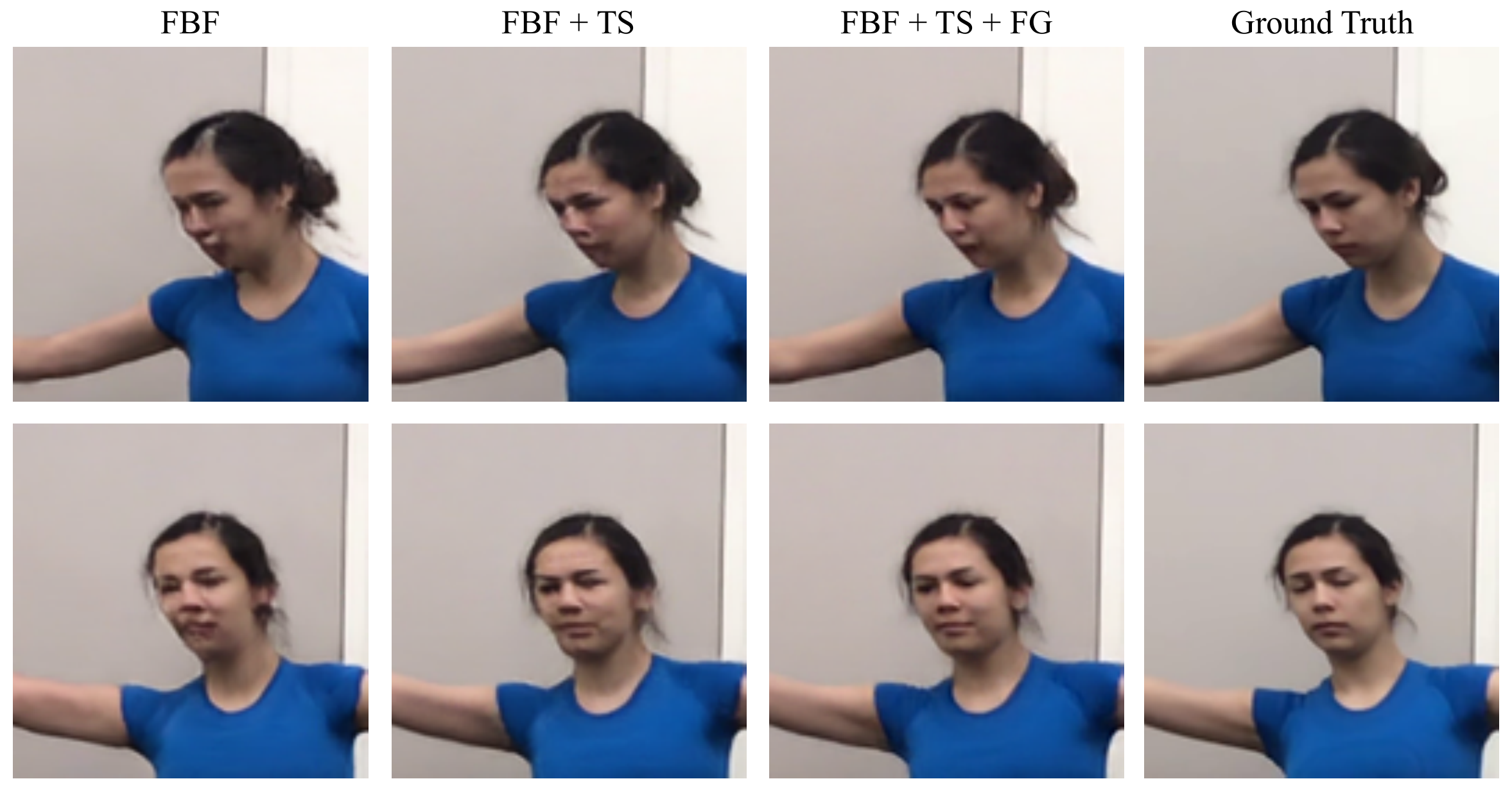}
  \caption{Face image comparison on held-out data. We compare frame-by-frame synthesis (FBF), adding temporal smoothing (FBF+TS) and our full model (FBF+TS+FG). }
  \label{fig:faceresults}
  \vspace{-0.18in}
\end{figure}

\begin{figure}
\centering
  \includegraphics[width=\linewidth]{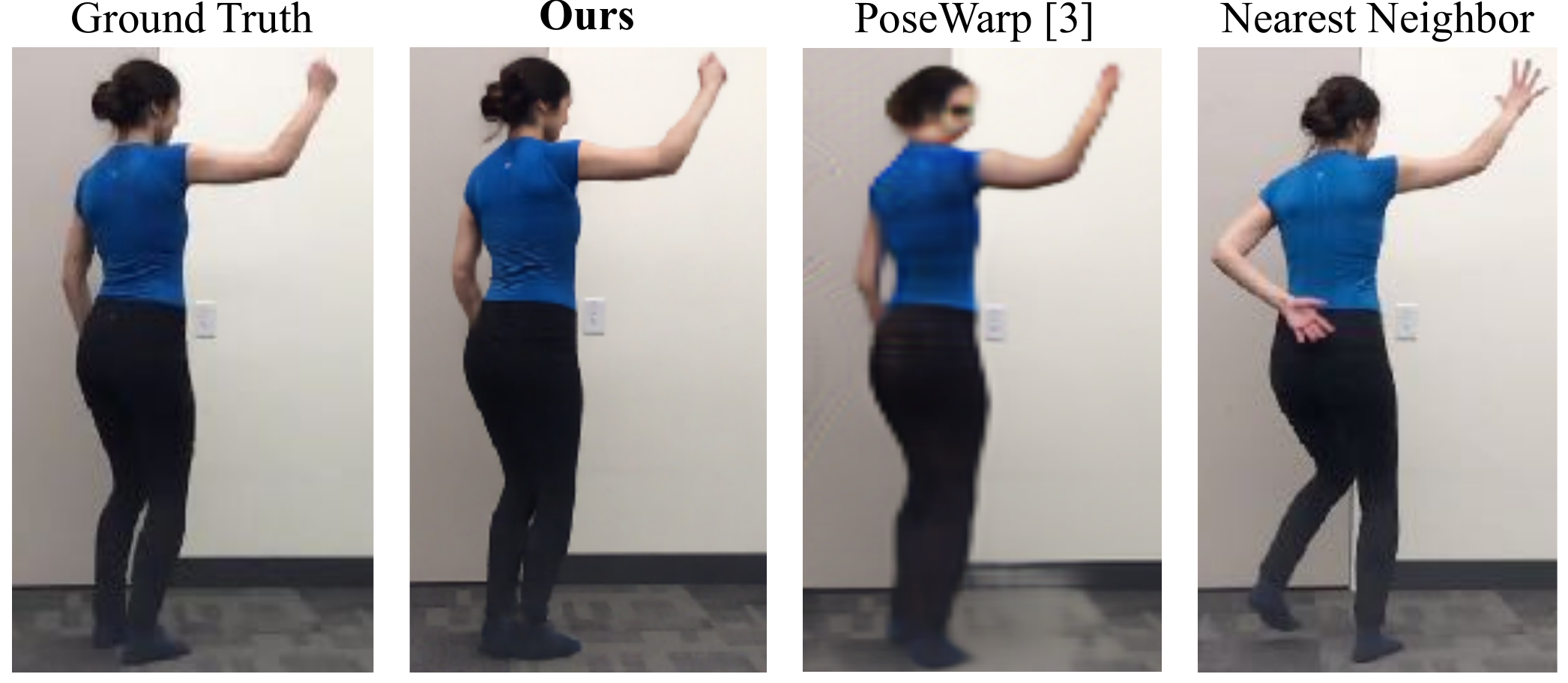}
  \caption{Comparison between our model,~\cite{balakrishnan2018synthesizing}, and nearest neighbors on single-frame synthesis on held-out data.}
  \vspace{-0.2in}
\label{fig:qualitative-baselines}
\end{figure}

Transfer results for multiple source and target subjects can be seen in Figure~\ref{fig:results}.
The advantage of using the Face GAN module can be seen in a single frame comparison in Figure~\ref{fig:faceresults}.
As mentioned,~\cite{balakrishnan2018synthesizing} is designed for single image synthesis. 
Nonetheless, even for a single frame transfer, we outperform~\cite{balakrishnan2018synthesizing} as we show in Figure~\ref{fig:qualitative-baselines}.

While the above single-image and quantitative results (Section~\ref{quantitative}) suggest the superiority of our approach, more significant difference can be observed in our video. There we find the temporal modeling produces more frame to frame coherence than the frame-by-frame ablation, and that adding a specialized facial generator and discriminator adds considerable detail and realism.

\begin{figure*}[!h]
\centering
  \includegraphics[width=0.99\linewidth]{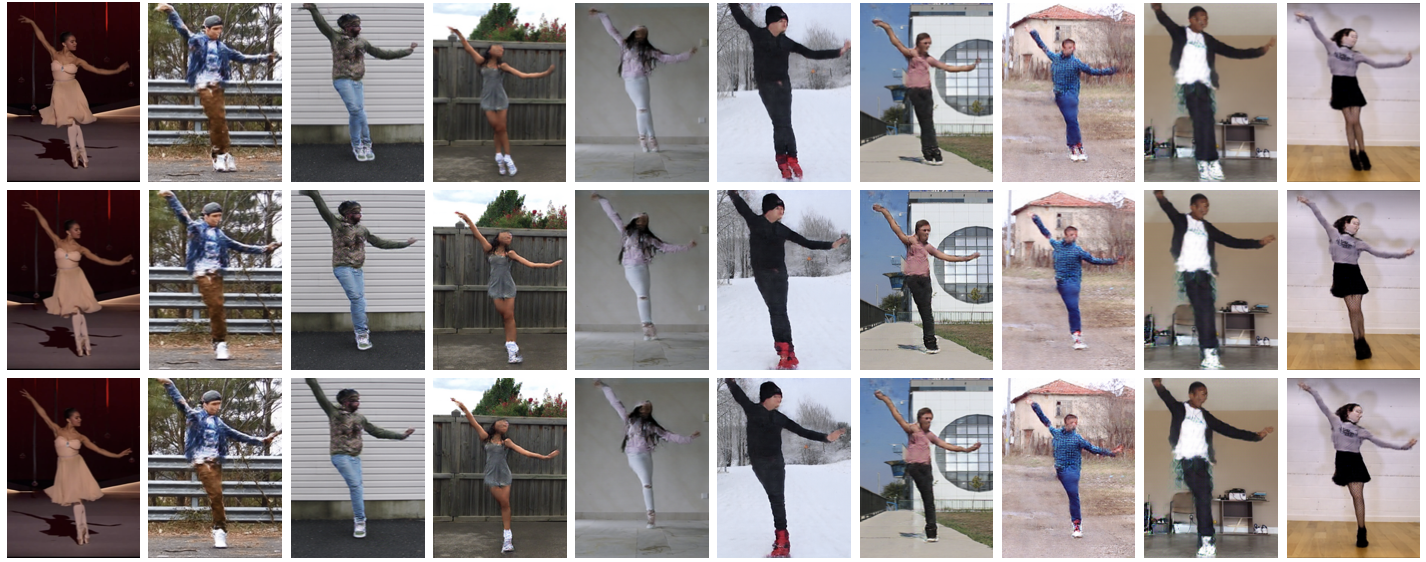}
    \vspace{-.1in}
    \caption{Multi-subject synchronized dancing. By applying the same source motion to multiple subjects, we can create the effect of them performing synchronized dance moves.
    }
  \label{fig:lotsofpeople}
  \vspace{-0.2in}
\end{figure*}

\section{Detecting Fake Videos}
\label{sec:fake}

Recent progress on image synthesis and generative models has narrowed the gap between synthesized and real images and videos, which has raised legal and ethical questions on video authenticity (among many other social implications). Given the high quality of our results, it is important to investigate mechanisms for detecting computer-generated videos including ones generated by our model.

We train a fake-detector to identify fake videos created by our system --- given a video, the fake-detector flags it as real or fake. We train the fake-detector in a parallel fashion to our synthesis process, to classify whether a sequence of $2$ consecutive frames is real (from ground-truth frames) or fake (from our generation). This allows the fake-detector to exploit cues based on the fidelity of individual frames as well as consistency across time. To make a decision for the whole video in question, we multiply the decision probabilities for all consecutive frame pairs. Details of the network architecture are included in Section~\ref{implementation}.
For the purpose of training the fake-detector, we collect a $62$-subject set of short $1920\times1080$ resolution dancing videos. This larger dataset is collected from public YouTube videos where a subject dances in front of a static camera for an average of $3$ minutes. We split this set into $48$ subjects for training and $14$ held-out subjects for testing.

We train a separate synthesis model for each of the $48$ train subjects to produce fake content for detection. By training our fake-detector on multiple fake videos depicting a large set of subjects we ensure that it generalizes to detecting fakes of different people and does not over-fit to one or two individuals. We note that since each person dancing performs a rich set of motions we require less training data than for detecting fakes in still images.

\begin{table}[t]
\centering
\begin{tabular}{lccc}
\multicolumn{1}{l}{Source Motion} & \multicolumn{1}{c}{Same subject} &  \multicolumn{1}{c}{Mars} &  \multicolumn{1}{c}{Copeland}\\ \midrule
Accuracy & 95.68\% & 96.70\% & 97.00\% \\
\end{tabular}
\caption{Fake detection average accuracy for held-out target subjects. As seen in the rows, fake videos were created for each target subject using same-subject and different-subject source motions.}
\vspace{-0.25in}
\label{table:fakes}
\end{table}

We evaluate our fake-detector on synthesized videos for $14$ held-out test subjects. We use both motion taken from the same subject (where the source and target are the same person) and motion driven by a different source subject (Bruno Mars and Misty Copeland) to synthesize fake videos for each held out subject. Our results are shown in Table~\ref{table:fakes}. Overall, the fake-detector successfully distinguishes real and fake sequences regardless of where the source motion is from. As expected, our fake detection accuracy is lowest for same-person motion transfer, and is highest for transfer of motion from a prima ballerina (Misty Copeland).

 \vspace{-2mm}
\section{Potential Applications}
One fun application of our system is to create a motion-synchronized dancing video with multiple subjects (say, for making a family reunion video). Given trained synthesis models for multiple subjects, we use the same source video to drive the motion of all target subjects --- creating an effect of them performing the same dance moves in a synchronized manner. See Figure~\ref{fig:lotsofpeople} and the video.

Several systems based on our prototype description were recently successfully employed commercially. One example is an augmented reality stage performance art piece where a 3D-rendered dancer appears to float next to a real dancer~\cite{mcdonald}. Another is an in-game entertainment application 
making NBA players dance~\cite{xpire}. 

\section{Limitations and Discussion}

Our relatively simple model is usually able to create arbitrarily long, good-quality  videos of a target person dancing given the movements of a source dancer to follow. However, it suffers from several limitations.

We have included examples of visual artifacts in Figure~\ref{fig:failure-cases}. On the left, our model struggles with loose clothing or hair which is not conveyed well through pose. The middle columns show a missing right arm which was not detected by OpenPose. On the right we observe some texture artifacts in shirt creases. Further work could focus on improving results by combining target videos with different clothing or scene lighting, improving pose detection systems, and mitigating the artifacts caused by high frequency textures in loose/wrinkled clothing or hair.

Our pose normalization solution does not account for different limb lengths or camera positions. These discrepancies additionally widen the gap between the motion seen in training and testing. However, our model is able to generalize to new motions fairly well from the training data. When filming a target sequence, we have no specific source motion in mind and do not require the target subject performing similar motions to any source. We instead learn a single model that generalizes to a wide range of source motion. However our model sometimes struggles to extrapolate to radically different poses. For example, artifacts can occur if the source motion contains extreme poses such as handstands if the target training data did not contain such upside-down poses. Future work could focus on the training data, i.e. what poses and how many are needed to learn a effective model. This area relates to work on understanding which training examples are most influential~\cite{koh2017understanding}.

\vspace{-5mm}
\paragraph{Acknowledgements}  We thank Andrew Owens for the catchy title.  This work was supported, in part, by NSF grant  IIS-1633310 and research gifts from Adobe, eBay, and Google. 

{\small
\bibliographystyle{ieee_fullname}
\bibliography{egbib}

\begin{thebibliography}{10}\itemsep=-1pt

\bibitem{aberman2019deep}
Kfir Aberman, Mingyi Shi, Jing Liao, D Liscbinski, Baoquan Chen, and Daniel
  Cohen-Or.
\newblock Deep video-based performance cloning.
\newblock In {\em Computer Graphics Forum}, volume~38, pages 219--233. Wiley
  Online Library, 2019.

\bibitem{baddar2017dynamics}
Wissam~J Baddar, Geonmo Gu, Sangmin Lee, and Yong~Man Ro.
\newblock Dynamics transfer gan: Generating video by transferring arbitrary
  temporal dynamics from a source video to a single target image.
\newblock {\em arXiv preprint arXiv:1712.03534}, 2017.

\bibitem{balakrishnan2018synthesizing}
Guha Balakrishnan, Amy Zhao, Adrian~V Dalca, Fredo Durand, and John Guttag.
\newblock Synthesizing images of humans in unseen poses.
\newblock In {\em CVPR}, 2018.

\bibitem{Recycle-GAN}
Aayush Bansal, Shugao Ma, Deva Ramanan, and Yaser Sheikh.
\newblock Recycle-gan: Unsupervised video retargeting.
\newblock In {\em ECCV}, 2018.

\bibitem{bregler1997video}
Christoph Bregler, Michele Covell, and Malcolm Slaney.
\newblock Video rewrite: Driving visual speech with audio.
\newblock In {\em Proceedings of the 24th annual conference on Computer
  graphics and interactive techniques}, pages 353--360. ACM
  Press/Addison-Wesley Publishing Co., 1997.

\bibitem{cao2017realtime}
Zhe Cao, Tomas Simon, Shih-En Wei, and Yaser Sheikh.
\newblock Realtime multi-person {2D} pose estimation using part affinity
  fields.
\newblock In {\em CVPR}, 2017.

\bibitem{casasEG2014}
Dan Casas, Marco Volino, John Collomosse, and Adrian Hilton.
\newblock {4D Video Textures for Interactive Character Appearance}.
\newblock {\em Computer Graphics Forum (Proceedings of EUROGRAPHICS)},
  33(2):371--380, 2014.

\bibitem{chen2017photographic}
Qifeng Chen and Vladlen Koltun.
\newblock Photographic image synthesis with cascaded refinement networks.
\newblock In {\em IEEE International Conference on Computer Vision (ICCV)},
  volume~1, page~3, 2017.

\bibitem{cheung2004markerless}
German~KM Cheung, Simon Baker, Jessica Hodgins, and Takeo Kanade.
\newblock Markerless human motion transfer.
\newblock In {\em 3D Data Processing, Visualization and Transmission, 2004.
  3DPVT 2004. Proceedings. 2nd International Symposium on}, pages 373--378.
  IEEE, 2004.

\bibitem{de2018semi}
Rodrigo de Bem, Arnab Ghosh, Thalaiyasingam Ajanthan, Ondrej Miksik, N
  Siddharth, and Philip Torr.
\newblock A semi-supervised deep generative model for human body analysis.
\newblock In {\em Proceedings of the European Conference on Computer Vision
  (ECCV)}, pages 0--0, 2018.

\bibitem{de2019conditional}
Rodrigo De~Bem, Arnab Ghosh, Adnane Boukhayma, Thalaiyasingam Ajanthan, N
  Siddharth, and Philip Torr.
\newblock A conditional deep generative model of people in natural images.
\newblock In {\em 2019 IEEE Winter Conference on Applications of Computer
  Vision (WACV)}, pages 1449--1458. IEEE, 2019.

\bibitem{Efros03}
Alexei~A. Efros, Alexander~C. Berg, Greg Mori, and Jitendra Malik.
\newblock Recognizing action at a distance.
\newblock In {\em IEEE International Conference on Computer Vision}, pages
  726--733, Nice, France, 2003.

\bibitem{vunet2018}
Patrick Esser, Ekaterina Sutter, and Bj{\"o}rn Ommer.
\newblock A variational u-net for conditional appearance and shape generation.
\newblock In {\em Proceedings of the IEEE Conference on Computer Vision and
  Pattern Recognition}, pages 8857--8866, 2018.

\bibitem{gleicher1998retargetting}
Michael Gleicher.
\newblock Retargetting motion to new characters.
\newblock In {\em Proceedings of the 25th annual conference on Computer
  graphics and interactive techniques}, pages 33--42. ACM, 1998.

\bibitem{hecker2008real}
Chris Hecker, Bernd Raabe, Ryan~W Enslow, John DeWeese, Jordan Maynard, and
  Kees van Prooijen.
\newblock Real-time motion retargeting to highly varied user-created
  morphologies.
\newblock In {\em ACM Transactions on Graphics (TOG)}, volume~27, page~27. ACM,
  2008.

\bibitem{isola2016image}
Phillip Isola, Jun-Yan Zhu, Tinghui Zhou, and Alexei~A Efros.
\newblock Image-to-image translation with conditional adversarial networks.
\newblock In {\em CVPR}, 2017.

\bibitem{Johnson2016Perceptual}
Justin Johnson, Alexandre Alahi, and Li Fei-Fei.
\newblock Perceptual losses for real-time style transfer and super-resolution.
\newblock In {\em European Conference on Computer Vision}, 2016.

\bibitem{joo2018generating}
Donggyu Joo, Doyeon Kim, and Junmo Kim.
\newblock Generating a fusion image: One's identity and another's shape.
\newblock In {\em Proceedings of the IEEE Conference on Computer Vision and
  Pattern Recognition}, pages 1635--1643, 2018.

\bibitem{kim2018DeepVideo}
Hyeongwoo Kim, Pablo Carrido, Ayush Tewari, Weipeng Xu, Justus Thies, Matthias
  Niessner, Patrick P{\'e}rez, Christian Richardt, Michael Zollh{\"o}fer, and
  Christian Theobalt.
\newblock Deep video portraits.
\newblock {\em ACM Transactions on Graphics (TOG)}, 37(4):163, 2018.

\bibitem{kim2017learning}
Taeksoo Kim, Moonsu Cha, Hyunsoo Kim, Jung~Kwon Lee, and Jiwon Kim.
\newblock Learning to discover cross-domain relations with generative
  adversarial networks.
\newblock In {\em Proceedings of the 34th International Conference on Machine
  Learning-Volume 70}, pages 1857--1865. JMLR. org, 2017.

\bibitem{koh2017understanding}
Pang~Wei Koh and Percy Liang.
\newblock Understanding black-box predictions via influence functions.
\newblock In {\em Proceedings of the 34th International Conference on Machine
  Learning-Volume 70}, pages 1885--1894. JMLR. org, 2017.

\bibitem{lassner2017generative}
Christoph Lassner, Gerard Pons-Moll, and Peter~V Gehler.
\newblock A generative model of people in clothing.
\newblock In {\em Proceedings of the IEEE International Conference on Computer
  Vision}, pages 853--862, 2017.

\bibitem{lee1999hierarchical}
Jehee Lee and Sung~Yong Shin.
\newblock A hierarchical approach to interactive motion editing for human-like
  figures.
\newblock In {\em Proceedings of the 26th annual conference on Computer
  graphics and interactive techniques}, pages 39--48. ACM Press/Addison-Wesley
  Publishing Co., 1999.

\bibitem{Liu2018Neural}
Lingjie Liu, Weipeng Xu, Michael Zollhoefer, Hyeongwoo Kim, Florian Bernard,
  Marc Habermann, Wenping Wang, and Christian Theobalt.
\newblock Neural rendering and reenactment of human actor videos.
\newblock {\em ACM Transactions on Graphics 2019 (TOG)}, 2019.

\bibitem{liu2017unsupervised}
Ming-Yu Liu, Thomas Breuel, and Jan Kautz.
\newblock Unsupervised image-to-image translation networks.
\newblock In {\em Advances in Neural Information Processing Systems}, pages
  700--708, 2017.

\bibitem{liu2016coupled}
Ming-Yu Liu and Oncel Tuzel.
\newblock Coupled generative adversarial networks.
\newblock In {\em Advances in neural information processing systems}, pages
  469--477, 2016.

\bibitem{ma2017pose}
Liqian Ma, Xu Jia, Qianru Sun, Bernt Schiele, Tinne Tuytelaars, and Luc
  Van~Gool.
\newblock Pose guided person image generation.
\newblock In {\em Advances in Neural Information Processing Systems}, pages
  406--416, 2017.

\bibitem{ma2018disentangled}
Liqian Ma, Qianru Sun, Stamatios Georgoulis, Luc Van~Gool, Bernt Schiele, and
  Mario Fritz.
\newblock Disentangled person image generation.
\newblock In {\em CVPR}, pages 99--108, 2018.

\bibitem{Martin-Brualla:2018:LEP:3272127.3275099}
Ricardo Martin-Brualla, Rohit Pandey, Shuoran Yang, Pavel Pidlypenskyi,
  Jonathan Taylor, Julien Valentin, Sameh Khamis, Philip Davidson, Anastasia
  Tkach, Peter Lincoln, Adarsh Kowdle, Christoph Rhemann, Dan~B Goldman, Cem
  Keskin, Steve Seitz, Shahram Izadi, and Sean Fanello.
\newblock Lookingood: Enhancing performance capture with real-time neural
  re-rendering.
\newblock {\em ACM Trans. Graph.}, 37(6):255:1--255:14, Dec. 2018.

\bibitem{mcdonald}
Kyle McDonald.
\newblock {Dance x Machine Learning: First Steps}.
\newblock \url{https://medium.com/@kcimc/discrete-figures-7d9e9c275c47}, 2019.
\newblock [Online; accessed 21-March-2019].

\bibitem{MoriBEEM04}
Greg Mori, Alex Berg, Alexei Efros, Ashley Eden, and Jitendra Malik.
\newblock Video based motion synthesis by splicing and morphing.
\newblock Technical Report UCB//CSD-04-1337, University of California,
  Berkeley, June 2004.

\bibitem{Guler2018DensePose}
Iasonas~Kokkinos R{\i}za Alp~G\"uler, Natalia~Neverova.
\newblock Densepose: Dense human pose estimation in the wild.
\newblock {\em arXiv}, 2018.

\bibitem{Siarohin_2018_CVPR}
Aliaksandr Siarohin, Enver Sangineto, Stéphane Lathuilière, and Nicu Sebe.
\newblock Deformable gans for pose-based human image generation.
\newblock In {\em The IEEE Conference on Computer Vision and Pattern
  Recognition (CVPR)}, June 2018.

\bibitem{simon2017hand}
Tomas Simon, Hanbyul Joo, Iain Matthews, and Yaser Sheikh.
\newblock Hand keypoint detection in single images using multiview
  bootstrapping.
\newblock In {\em CVPR}, 2017.

\bibitem{simonyan2014very}
Karen Simonyan and Andrew Zisserman.
\newblock Very deep convolutional networks for large-scale image recognition.
\newblock {\em arXiv preprint arXiv:1409.1556}, 2014.

\bibitem{tulyakov2017mocogan}
Sergey Tulyakov, Ming-Yu Liu, Xiaodong Yang, and Jan Kautz.
\newblock Mocogan: Decomposing motion and content for video generation.
\newblock {\em IEEE Conference on Computer Vision and Pattern Recognition
  (CVPR)}, 2018.

\bibitem{villegas2018neural}
Ruben Villegas, Jimei Yang, Duygu Ceylan, and Honglak Lee.
\newblock Neural kinematic networks for unsupervised motion retargetting.
\newblock In {\em The IEEE Conference on Computer Vision and Pattern
  Recognition (CVPR)}, June 2018.

\bibitem{villegas2017learning}
Ruben Villegas, Jimei Yang, Yuliang Zou, Sungryull Sohn, Xunyu Lin, and Honglak
  Lee.
\newblock Learning to generate long-term future via hierarchical prediction.
\newblock {\em arXiv preprint arXiv:1704.05831}, 2017.

\bibitem{pos_iccv2017}
Jacob Walker, Kenneth Marino, Abhinav Gupta, and Martial Hebert.
\newblock The pose knows: Video forecasting by generating pose futures.
\newblock In {\em International Conference on Computer Vision}, 2017.

\bibitem{wang2018vid2vid}
Ting-Chun Wang, Ming-Yu Liu, Jun-Yan Zhu, Guilin Liu, Andrew Tao, Jan Kautz,
  and Bryan Catanzaro.
\newblock Video-to-video synthesis.
\newblock In {\em Advances in Neural Information Processing Systems (NeurIPS)},
  2018.

\bibitem{wang2017highres}
Ting-Chun Wang, Ming-Yu Liu, Jun-Yan Zhu, Andrew Tao, Jan Kautz, and Bryan
  Catanzaro.
\newblock High-resolution image synthesis and semantic manipulation with
  conditional gans.
\newblock In {\em Proceedings of the IEEE Conference on Computer Vision and
  Pattern Recognition}, 2018.

\bibitem{wang2004image}
Zhou Wang, Alan~C Bovik, Hamid~R Sheikh, and Eero~P Simoncelli.
\newblock Image quality assessment: from error visibility to structural
  similarity.
\newblock {\em IEEE transactions on image processing}, 13(4):600--612, 2004.

\bibitem{wei2016cpm}
Shih-En Wei, Varun Ramakrishna, Takeo Kanade, and Yaser Sheikh.
\newblock Convolutional pose machines.
\newblock In {\em CVPR}, 2016.

\bibitem{xpire}
Xpire.
\newblock {Using AI to make NBA players dance}.
\newblock \url{https://tinyurl.com/y3bdj5p5}, 2019.
\newblock [Online; accessed 21-March-2019].

\bibitem{xu2011video}
Feng Xu, Yebin Liu, Carsten Stoll, James Tompkin, Gaurav Bharaj, Qionghai Dai,
  Hans-Peter Seidel, Jan Kautz, and Christian Theobalt.
\newblock Video-based characters: creating new human performances from a
  multi-view video database.
\newblock In {\em ACM Transactions on Graphics (TOG)}, volume~30, page~32. ACM,
  2011.

\bibitem{zanfir2018human}
Mihai Zanfir, Alin-Ionut Popa, Andrei Zanfir, and Cristian Sminchisescu.
\newblock Human appearance transfer.
\newblock In {\em Proceedings of the IEEE Conference on Computer Vision and
  Pattern Recognition}, pages 5391--5399, 2018.

\bibitem{zhang2018perceptual}
Richard Zhang, Phillip Isola, Alexei~A Efros, Eli Shechtman, and Oliver Wang.
\newblock The unreasonable effectiveness of deep features as a perceptual
  metric.
\newblock In {\em CVPR}, 2018.

\bibitem{CycleGAN2017}
Jun-Yan Zhu, Taesung Park, Phillip Isola, and Alexei~A Efros.
\newblock Unpaired image-to-image translation using cycle-consistent
  adversarial networkss.
\newblock In {\em Computer Vision (ICCV), 2017 IEEE International Conference
  on}, 2017.

\end{thebibliography}
}

\section{Appendix} \label{appendix}

\subsection{Video Demonstration}
Our video demo can be found at \url{https://youtu.be/mSaIrz8lM1U} and examples from our comparison to baselines and ablation study can be found at \url{https://youtu.be/sQD0WVS0blg}.

\subsection{Implementation Details} \label{implementation}
Our generator and discriminator architectures are modified from pix2pixHD [41] to handle the temporal setting. We follow the progressive learning schedule from pix2pixHD and learn to synthesize at $512 \times 256$ at the first (global) stage, and then upsample to $1024 \times 512$ at the second (local) stage. For predicting face residuals, we use the global generator of pix2pixHD and a single $70\times70$ Patch-GAN discriminator [16]. We set hyperparameters $\lambda_{P} = 5$ and $\lambda_{VGG} = 10$ during the global and local training stages respectively. For the dataset collected in Section~\ref{sec:setup}, we trained the global stage for 5 epochs, the local stage for 30 epochs, and the face GAN for 5 epochs.

For the perceptual loss $\mathcal{L}_{P}$, we compare the \texttt{conv1\_1}, \texttt{conv2\_1}, \texttt{conv3\_1}, \texttt{conv4\_1}, and \texttt{conv5\_1} layer outputs of the VGG-19 network.

Our generator and discriminator architectures follow that presented by Wang et al. [41]. The fake-detector architectures matches that of the discriminator with a final fully connected layer.

\subsection{Dataset Collection} \label{collection}
Our dataset of long target videos consists of footage we filmed ourselves from $8$ to $17$ minutes with $4$ videos at $1920\times1080$ resolution and $1$ at $1280\times720$.
Our goal in collecting a dataset of target videos is to provide the community with open-source data for which we explicitly collect release forms in which subjects allow their data to be released to other researchers. We recruited target subjects from different sources: friends, professional dancers, reporters etc. To learn the appearance of the target subject in many poses, it is important that the target video captures a sufficient range of motion and sharp frames with minimal blur. Similarly, we used a stationary camera to ensure a static background in all frames. To ensure the quality of the frames, we filmed our target subjects for between $8$ and $30$ minutes of real time footage at $120$ frames per second using a modern cellphone camera, and use the first $20\%$ of the footage for training and the last $80\%$ for testing. Since our pose representation does not encode information about clothes and hair, we instructed our target subjects not to wear loose clothing and to tie up long hair.

In contrast, source videos can be easily collected online as we only require decent pose detections on these. We therefore use in-the-wild single-dancer videos where the only restriction we enforce is a static camera position.

\subsection{Comparison with vid2vid}

\begin{figure}
\centering
  \includegraphics[width=\linewidth]{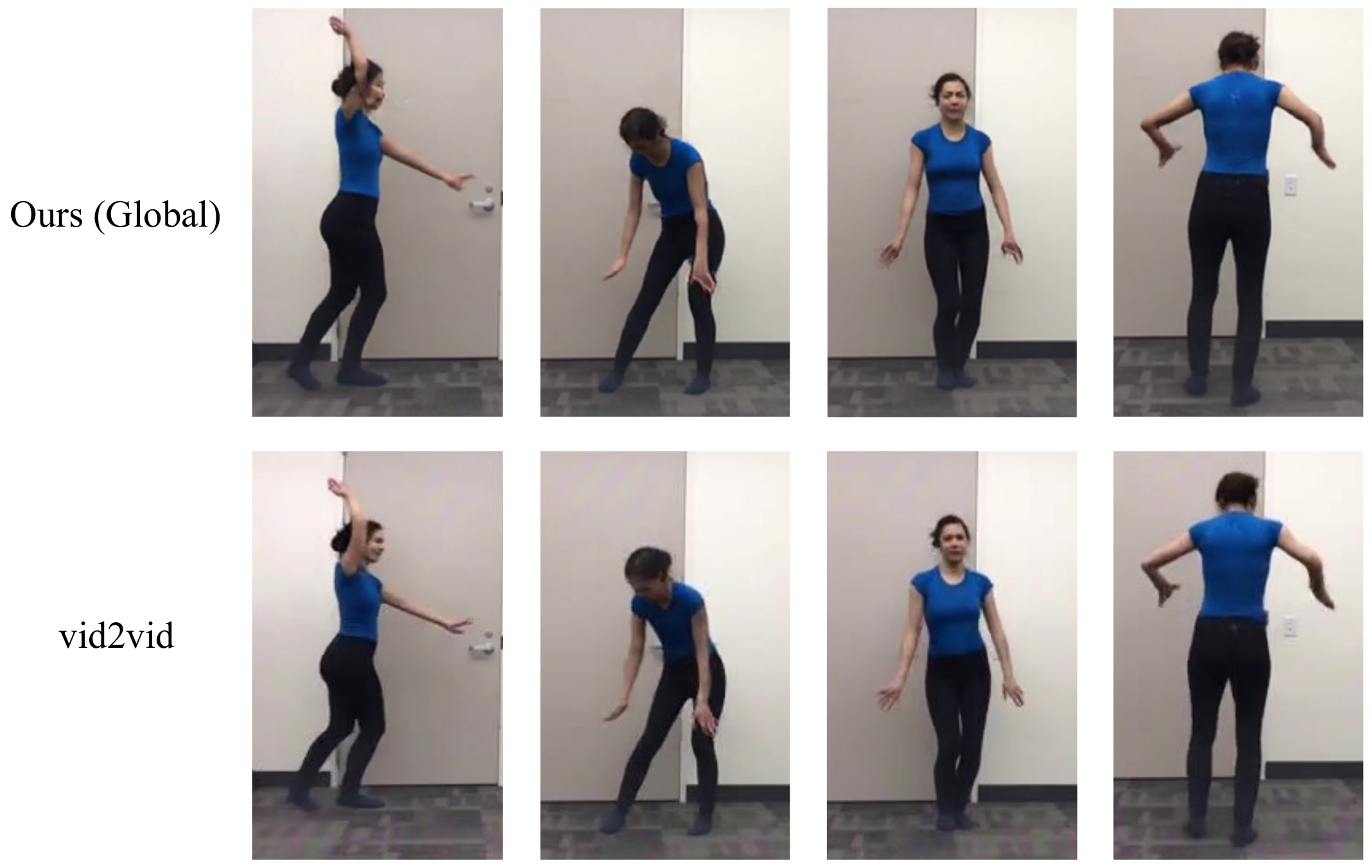}
  \caption{We compare a lower resolution version of our model without a Face GAN (top) with a lower resolution of vid2vid [41]. We find our results comparable. }
  \vspace{-0.2in}
\label{fig:vid2vid}
\end{figure}

We also compare our model with a concurrent video synthesis framework called vid2vid~\cite{wang2018vid2vid}. The excessive requirement of memory and computing power of vid2vid prohibits us from comparing with their model in the high resolution setup. Instead, we train both our model and theirs in lower resolution ($512 \times 256$). Our system and vid2vid generally perform similarly and produce results of comparable quality. We provide a qualitative comparison in Figure~\ref{fig:vid2vid}.

\subsection{Global Pose Normalization Details} \label{posenorm}

In this section we describe our normalization method to match poses between the source and target. Consider a case where the source subject is significantly taller in frame than the target or is slightly elevated above the target subject's in frame position. If we directly input the unmodified poses to our system, we may generate images of the target person which are not congruent with the scene. In this example, the target person may appear large with respect to the background or surrounding objects, and may appear to be levitating since the input pose places the feet above the floor. Additionally, when generating an image from a very different pose from the in proportion and reasonably positioned poses in training, the overall quality of synthesis is expected to decline. Therefore we design a method to reasonably match the poses by finding a suitable transformation between the source and target poses. We parametrize this transformation in terms of a scale and translation factor applied to all pose keypoints for a given frame.

To find a suitable translation factor, we need to determine the position of both subjects within their respective frames. We first find the closest position $s_{close}$ and farthest position $s_{far}$ the source subject is away from the camera in their video. Similarly, we do the same for the target by determining $t_{close}$ and $t_{far}$ respectively. The goal is then to map the close and far range of the source to that of the target subject as to match the positions of both subjects, i.e. $s_{far} \mapsto t_{far}$ and $s_{close} \mapsto t_{close}$. Given a frame where the source is at position $y$, we then translate the source's pose vertically by:
\begin{equation}
translation = t_{far} + \frac{y - s_{far}}{s_{close} - s_{far}} (t_{close} - t_{far})
\end{equation}
In practice, we use the average of the y coordinates of the subject's ankles to determine the position within a given frame.

To reasonably scale the source poses, we determine the heights of each subject at their closest and farthest positions in their video - denote these quantities as $h_{s_{close}}, h_{s_{far}}$ for the source and $h_{t_{close}}, h_{t_{far}}$ for the target subjects respectively. We then determine separate scales for the close position given by $c_{close} = \frac{h_{t_{close}}}{h_{s_{close}}}$ and similarly for the far position given by $c_{far} = \frac{h_{t_{far}}}{h_{s_{far}}}$. When given a frame where the source is at position $y$, we scale the source's pose (in both x, y directions) by:
\begin{equation}
scale =  c_{far} + \frac{y - s_{far}}{s_{close} - s_{far}} (c_{close} - c _{far})
\end{equation}
We use the euclidean distance between the average ankle position and the nose keypoint of our given pose as the subject's height in a given frame.

After the translation and scale factors have been determined for a given source pose, we then add the translation to all keypoints and then apply the scale factor so that the ankle $y$ positions remain the same (i.e. the ground is the x axis).

Given poses from a subject, we find the close position by taking the maximum y coordinate of their average ankle position over all frames.
$$s_{close} = \max{\{ \frac{s_{ankle1} + s_{ankle2}}{2} \} }$$
The far position is found by clustering the y ankle coordinates which are less than (or spatially above) the median ankle position and about the same distance as the maximum ankle position's distance to the median ankle position. If we denote $S =\frac{s_{ankle1} + s_{ankle2}}{2}$ as the average ankle position in a given frame, then the clustering is as described by the set 
\begin{equation}
\max \{S : ||S - s_{med}| < \alpha |s_{close} - s_{med}||  \} \cap \{S < s_{med} \}
\end{equation}
where $s_{med}$ is the median foot position, $max$ is the maximum ankle position, and $\epsilon$ and $\alpha$ are scalars. In practice we find setting $\alpha = 0.7$ generally works well, although this scalar can be finetuned on a case by case basis since it depends highly on the camera height and the subject's range of motion.

\end{document}